\documentclass{kluwer}
\newdisplay{guess}{Conjecture}
\begin{document}
\begin{article}
\begin{opening}
\tolerance=10000 \hbadness=10000
\def\verbfont{\small\tt} 
\def\note #1]{{\bf #1]}} 
\def\be{\begin{equation}}
\def\ee{\end{equation}} 
\def\bearr{\begin{eqnarray}} 
\def\eearr{\end{eqnarray}}
\def\barr{\begin{array}}
\title{HOW GOOD ARE THE PREDICTIONS FOR OSCILLATION FREQUENCIES?}
\author{KIRAN \surname{JAIN}}
\author{S. C. \surname{TRIPATHY}}
\author{A. \surname{BHATNAGAR}}
\runningauthor{JAIN, TRIPATHY, AND BHATNAGAR}
\runningtitle{PREDICTIONS FOR OSCILLATION FREQUENCIES}
\institute{Udaipur Solar Observatory, A unit of Physical Research Laboratory, Off Bari Road, 
Dewali, P.  B.  No.  198, Udaipur 313001, India 
%(E-mails:  kiran@uso.ernet.in, arvind@uso.ernet.in 
} 
\date{}
\begin{abstract}
We have used available intermediate degree {\it p}-mode frequencies for the solar 
cycle 23 to check the validity of previously derived empirical relations for 
frequency shifts 
(Jain {\it et al.}: 2000, {\it Solar Phys.} {\bf 192}, 487). We find that 
 the calculated and observed frequency shifts during the rising phase of
the cycle 23 are in good agreement. The observed frequency shift 
from minimum to maximum of this cycle as calculated from MDI frequency 
data sets 
is 251 $\pm$ 7 nHz and from GONG data is 238 $\pm$ 11 nHz. These values are in close agreement 
with the empirically predicted value 
of 271 $\pm$ 22 nHz.

\end{abstract}

\end{opening}

\section{Introduction} 

An understanding of the physical processes responsible for the changes in the solar 
{\it p}-mode frequencies could provide an important clue to the inner workings of 
the solar activity cycle. It is now well established that these frequencies change with time
(Woodard and Noyes, 1985) and show a positive correlation with activity indices 
(Woodard {\it et al.}, 1991;
Bachmann and Brown, 1993). Over the last two decades, attempts have been made
to precisely measure  the  changes in {\it p}-mode oscillation frequencies.
More recently, with the Global Oscillation Network Group (GONG) (Harvey {\it et
al.}, 1996)  and Michelson Doppler Imager (MDI)   on board Solar and
Heliospheric Observatory (Scherrer {\it et al.}, 1995)   instruments, the
measurements are made   consistently with an accuracy of one part in 10$^5$ or
better. These continuous   data sets further confirm that the oscillation
frequencies are well correlated with  the activity indices (Jain, Tripathy,
and Bhatnagar, 2000 and {\it references therein}).  Following a different
approach and using a data set of eight years  between 1981 and 1989, Rhodes
{\it et al.} (1993) reported that the frequency shifts are also correlated
with the change in various activity  indicators. This study was  extended
to the rising part of the cycle 23   by Jain {\it et al.} (2000; hereafter
JTBK).  Using the GONG frequencies for the period May 1995 to October 1998,
they  confirmed that the frequency shifts are better correlated with the
change in  activity  indices. In an attempt to quantify the changes in mode
frequencies, JTBK    derived  empirical relations between the shift  in
frequencies and change in the level of activity indices and  showed  that
these relations do not change significantly from cycle to cycle.   Using a
limited data set  from GONG network for the ascending part of the current
cycle 23,  it was found that the  calculated and observed frequency
shifts were in close agreement.

The motivation of this paper is to check the validity of the derived 
relations for the maximum and descending phase of the  current solar cycle.
%The motivation of this paper is to check the validity of the derived relations for the
%current solar cycle  including the maximum  phase and to extend to the
%declining phase of this cycle from the predicted value of solar activity index.
This is accomplished by determining the frequency shifts using the smoothed
sunspot number and equation (8) of JTBK.  The calculated shifts are compared 
with the measured shifts obtained from  recent observations. We find that
the calculated and observed frequency shifts are  in close agreement
confirming  that the derived relations can be reliably used to estimate the
{\it p}-mode frequencies for past, present and future solar activity cycles,
if the solar acitivity index is  known.

\section{Frequency Data Sets}
In order to check the validity of emprical relations of JTBK, we have used intermediate 
degree {\it p}-mode frequencies for the solar cycle 23  from GONG and MDI instruments 
covering a period of more than five years. The GONG data consist of fifty three, 
108 day overlapping data 
sets starting from 7 May 1995 to 6 October, 
2000. This covers the declining phase of the solar cycle 22 and the rising phase of the
 solar cycle 23.  The MDI data consist of twenty four, 72 day data sets and
cover the period from 1 May, 1996 to 15 June, 2001 with two breakdowns in between. Since 
empirical relations of JTBK  are valid in the  spherical harmonic degree 
range 5 $\le$ $\ell$ $\le$ 99 and 
frequency range 1500~$\mu$Hz $\le$ $\nu$ $\le$ 3500 $\mu$Hz, we have considered only
those common modes which lie in between these two ranges. This selection criterion generated 
a total number of 327 common modes for GONG data and 651 modes for MDI data.   
	
\begin{figure}
\begin{center}
\leavevmode
\input epsf
\epsfxsize=3.75in \epsfbox{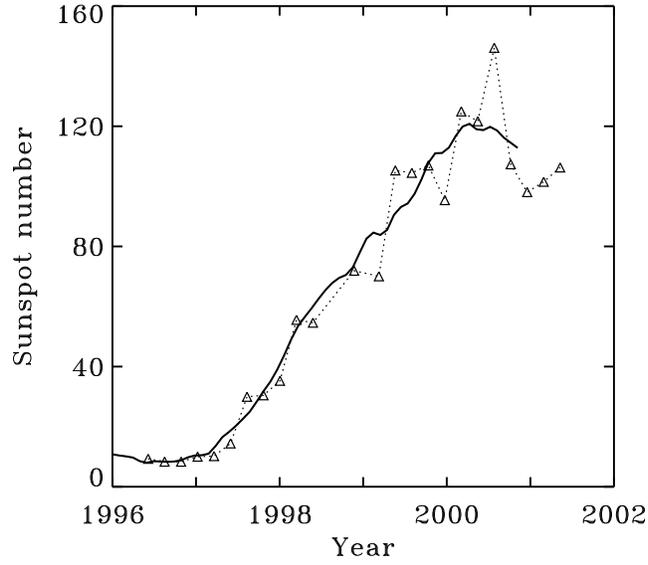}\\
\leavevmode
\caption{The smoothed sunspot number (solid line) and mean international sunspot number (triangles) 
averaged over each of twenty-four, 72-d observing periods of MDI. }
\end{center}
\end{figure} 

 \begin{figure}
\begin{center}
\leavevmode
\input epsf
\epsfxsize=4.0in \epsfbox{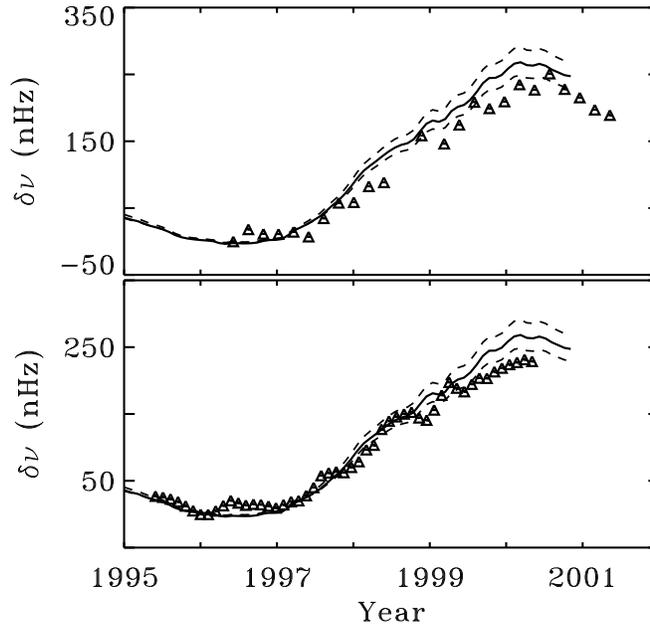}\\
\leavevmode
\caption{The calculated (solid line) and observed (triangles) frequency shifts for 1995-2000. 
The lower panel shows the variation in GONG frequencies and upper panel for MDI frequencies. 
The 1 $\sigma$ errors are shown by dashed lines.  }
\end{center}
\end{figure}

\section{Results and Discussion}

	We estimate the change in {\it p}-mode frequencies for solar cycle 23
using the equation (8) of JTBK:

\begin{eqnarray}
\delta\nu & =& (2.41 \pm 0.19)~\delta R_s - (0.48 \pm 1.68),
\end{eqnarray}
where $\delta \nu$ is given in nHz and  $\delta R_s$ is the 
change in smoothed sunspot number. The variation of sunspot number taken from the {\it Solar 
Geophysical Data} web page for the observing period
of MDI is shown in Figure 1. The estimated frequency shifts obtained from
Equation (1) are  plotted in Figure 2 for the period from  January, 1995 to
December, 2000. The measured shifts for both GONG (lower panel) and MDI (upper panel) 
data sets are also plotted in the same figure and  shows 
 that the observed  frequency shifts  are in  close agreement
 with those obtained from the empirical relation. During the 
 ascending phase, the GONG frequency 
 shifts agree better with the derived shifts while the frequency shifts of MDI data 
 sets are in close agreement during the maximum phase of the cycle.  
On an average,  we find that the deviation between the calculated and observed frequency shifts 
is within 1$\sigma$  error level. 

%agreement is
%better during the rising phase    of the solar cycle and deviation occurs
%around the maximum phase of the cycle.     

\begin{figure}
\begin{center}
\leavevmode
\input epsf
\epsfxsize=3.75in \epsfbox{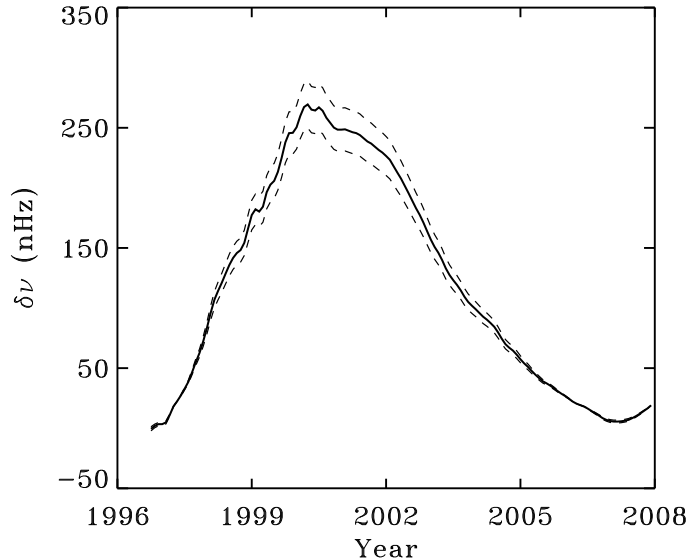}\\
\leavevmode
\caption{The estimated frequency shifts for the solar cycle 23 using predicted sunspot numbers. The
dashed lines represent the  errors in the calculation due to the errors in predicted activity index. }
\end{center}
\end{figure}

  Quantitatively, the derived shift
between the minimum and maximum of the current solar cycle amounts to 271
$\pm$ 22 nHz corresponding to the  maximum smoothed sunspot number of  120.8.
This can be compared to the observed  shifts of   
251 $\pm$ 7 nHz for MDI and 238 $\pm$ 11 nHz for GONG data and clearly shows a 
discrepancy near the maximum phase of the cycle indicating to the  
complex nature of the relationship that may exist  between activity index and the
frequecy shift. Earlier, JTBK had quoted a maximum 
shift of 265 $\pm$ 90 nHz corresponding to the predicted maximum
 sunspot number of 118 $\pm$ 35. 

 The estimated frequency shifts for the complete solar cycle 23 (1996 - 2008) 
are plotted in Figure~3. The solid line represents the estimated
 shifts calculated using the predicted smoothed sunspot number taken from the {\it Solar Geophysical
 Data} web page and dashed lines are the predicted 1$\sigma$ error.

  In summary, we have used available intermediate degree {\it p}-mode frequencies 
  from GONG and MDI projects for the solar 
cycle 23 to check the validity of previously derived empirical relations for 
frequency shifts 
(Jain {\it et al.}, 2000). We find that 
 the calculated frequency shifts  are in close agreement with the observed shifts during 
 the period considered in this analysis which includes the rising phase of the cycle 23.  
We conclude that the empirical relations as derivd by JTBK can be considered
as good predictors of frequency shifts for 
solar activity cycles. 
%The discepancy noticed around the maximum pahse of the solar cycle 
%indicates that the relationsh%ip between activity indices and the frequency 
%shift may be more complex.

\acknowledgements
 
This work utilises data from the Michelson Doppler Imager
 on the Solar and Heliospheric Observatory. SOHO is a mission of international cooperation
 between ESA and NASA. We thank J. Schou for providing us 
the MDI data. This work also utilizes data obtained by the Global Oscillation Network
Group (GONG) project, managed by the National Solar Observatory, a
Division of the National Optical Astronomy Observatories, which is
operated by AURA, Inc. under a cooperative agreement with the National
Science Foundation. The data were acquired by instruments operated by
the Big Bear Solar Observatory, High Altitude Observatory, Learmonth Solar 
Observatory, Udaipur Solar Observatory, Instituto de Astrophsico de 
Canaris, and Cerro Tololo Interamerican Observatory. This work is partially 
supported under the CSIR Emeritus Scientist Scheme and Indo-US collaborative 
programme $-$ NSF Grant INT-9710279.

\end{article}


\begin{thebibliography}{}
\bibitem{bach:93} Bachmann, K. T. and Brown, T. M.: 1993,  
{\it Astrophys. J.} {\bf 411}, L45.
\bibitem{harv:96} Harvey, J.W. {\it et al.}: 1996,  
{\it Science} {\bf 272}, 1284.
\bibitem{ksa:00} Jain, K., Tripathy, S. C. and Bhatnagar, A.: 2000, 
{\it Astrophys. J.} {\bf 542}, 521. 
\bibitem{jain:00} Jain, K., Tripathy, S. C., Bhatnagar, A. and Kumar, B.: 2000, 
{\it Solar Phys.} {\bf 192}, 487. 
\bibitem{rhodes:93} Rhodes, E. J., Jr., Cacciani, A., 
Korzennik, S. G., and Ulrich, R. K.: 1993, {\it Astrophys. J.} {\bf 406}, 714.
\bibitem{schr:95} Scherrer, P. H. {\it et al.}: 1995, 
{\it Solar Phys.} {\bf 162}, 129. 
\bibitem{Wood:85} Woodard, M.F. and Noyes, R.W.: 1985,{\it Nature} {\bf 318}, 449.
\bibitem{wood:91} Woodard, M. F., Kuhn, J. R., Murray, N., and Libbrecht, K. G.: 1991, 
{\it Astrophys. J.} {\bf 373}, L81.
\end{thebibliography}
\end{document}